\documentclass[11pt,a4paper]{article}

\pdfoutput=1 

\usepackage{jcappub} 
\usepackage[T1]{fontenc} 
\usepackage{amsmath,amssymb}
\usepackage{epsfig,graphicx}
\usepackage{subfigure}
\usepackage{graphicx}
\usepackage{soul}
\usepackage{hyperref}
\usepackage{rotating}
\usepackage{cancel}
\usepackage{bm}
\usepackage{xcolor}
\usepackage[font={small}]{caption}
\usepackage{comment}
\usepackage{cite}
\usepackage{psfrag}
\usepackage[numbers,sort&compress]{natbib}
\usepackage{color}

\newcommand{\R}{\chi}

\def\beq{\begin{equation}}
\def\eeq{\end{equation}}

\title{Relaxion Dark Matter from Stochastic Misalignment}

\author[a]{Aleksandr Chatrchyan}
\author[a,b]{and G\'{e}raldine Servant}

\affiliation[a]{Deutsches Elektronen-Synchrotron DESY, Notkestr.~85, 22607 Hamburg, Germany}
\affiliation[b]{II. Institute of Theoretical Physics, Universit\"{a}t Hamburg D-22761, Germany}

\emailAdd{aleksandr.chatrchyan@desy.de}
\emailAdd{geraldine.servant@desy.de}

\abstract{ Cosmological relaxation of the electroweak scale via Higgs-axion interplay, named as relaxion mechanism,  provides a dynamical solution to the Higgs mass hierarchy. In the original  proposal by Graham, Kaplan and Rajendran, the relaxion abundance today is too small to explain the dark matter of the universe because of the high suppression of the misalignment angle after inflation. It was then realised by Banerjee, Kim and Perez that reheating effects can displace the relaxion, thus enabling it to account for the dark matter abundance from the misalignment mechanism. However, this scenario is realised in a limited region of parameter space to avoid runaway. We show that in the regime where inflationary fluctuations dominate over the classical slow-roll, the ``stochastic misalignment" of the field due to fluctuations can be large. We study the evolution of the relaxion after inflation, including the high-temperature scenario, in which the barriers of the potential shrink and destabilise temporarily the local minimum. We open new regions of parameter space where the relaxion can naturally explain the observed dark matter density in the universe, towards larger coupling, larger  mass, larger mixing angle, smaller decay constant, as well as larger scale of inflation. }

\begin{document}

\begin{flushright}
	\footnotesize
	DESY-22-177 \\
\end{flushright}
\color{black}

\maketitle
\flushbottom

\section{Introduction}

A different approach from the traditional methods to address the Higgs hierarchy problem has been proposed lately, that relies on a Higgs-axion cosmological interplay \cite{Graham:2015cka}. The axion field introduced in this context is called the {\it relaxion},
and its role is to dynamically select a small value for the Higgs mass in the early universe. The potential for the coupled relaxion-Higgs system has the simple form
\beq
\label{eq_potential}
V(h,\phi) = -g\Lambda^3\phi + \frac{1}{2}[\Lambda^2 - g'\Lambda \phi]h^2 + \frac{\lambda_h}{4!}h^4 + \Lambda_{b}^4 \Bigl[1-\cos\Bigl(\frac{\phi}{f}\Bigr)\Bigr],
\eeq
where $\Lambda$ is the cut-off scale of the standard model (SM) Higgs effective field theory, $g$ and $g'$ are small dimensionless parameters that characterize the rolling potential and the relaxion-Higgs coupling, $\lambda_h$ is the quartic coupling of the Higgs, $f$ is the decay constant of the relaxion and $\Lambda_b$ describes the barriers of the relaxion potential. The height of the barriers $\Lambda_b$ is sensitive to the Higgs vev and vanishes if $\langle h \rangle =0$. As a consequence, starting from a large and positive Higgs mass squared, the slow-roll dynamics of the relaxion eventually brings it to a local minimum corresponding to a small and negative Higgs mass squared.

In a large part of the parameter region, the relaxion is light and stable on cosmological time scales. Such scalars are known to be generally good candidates for dark matter (DM) when produced via the misalignment mechanism~\cite{Preskill:1982cy, Abbott:1982af, Dine:1982ah, Arias:2012az}. In the original proposal \cite{Graham:2015cka}, the relaxion cannot address both the hierarchy problem and the DM. In this work, we re-visit the possibility that the relaxion solves the hierarchy problem and, at the same time, explains the observed DM density in the universe.

A long period of inflation is usually required for the relaxion to scan the Higgs mass and roll down to the correct local minimum (for a status review of alternative friction mechanisms, see~\cite{Fonseca:2019lmc}). The field is also subject to quantum fluctuations which, in the superhorizon limit, are determined by the inflationary Hubble scale $H_I$. These fluctuations produce a random walk for the relaxion, preventing it from stopping exactly at the minimum. On timescales much shorter compared to the total relaxation time, a (meta)equilibrium is established in the local minimum, of the form~\cite{Starobinsky:1994bd}
\beq
\label{eq_eq}
\rho(\phi) \propto \exp\Bigl( - \frac{8\pi^2V(\phi)}{3H_I^4} \Bigr),
\eeq
where $\rho(\phi)$ denotes the probability distribution for the field to have an average value $\phi$ inside a Hubble patch. The local minimum is expected to be long-lived, hence, the distribution is concentrated near the quadratic minimum of the potential. In this limit, it is approximately gaussian with a variance given by
\beq
\label{stochastic_variance}
\sigma_{\phi}^2 = \frac{3H_I^4}{8\pi^2m_{\phi}^2}.
\eeq
This ``stochastic'' misalignment is later converted into coherent oscillations of the field which can behave as DM. Such a stochastic window for standard QCD axion DM was investigated in detail in~\cite{Graham:2018jyp, Takahashi:2018tdu}.

The parameter region of the relaxion can be split into two parts. 
\begin{itemize}
	\item 
	$\boxed{H_I^3<g\Lambda^3}$\\
	In the so-called classical-beats-quantum (CbQ) regime, 
	the slow-roll of the field per Hubble time dominates over its random motion, as the field rolls down the potential. Most of the studies, including~\cite{Graham:2015cka}, considered the mechanism in this regime. Unfortunately, the stochastic misalignment can explain only a tiny fraction of the DM abundance in this case.
	
	\item
	$\boxed{H_I^3>g\Lambda^3}$\\
	Somewhat less explored is the quantum-beats-classical (QbC) regime, where the random walk of the relaxion dominates over its classical motion. The mechanism in this regime was investigated in our recent work~\cite{Chatrchyan:2022pcb}, as well as in~\cite{Nelson:2017cfv, Gupta:2018wif}. Using the Fokker-Planck equation it was shown that also in this case the field can roll down to the minimum with a small Higgs mass, successfully generating the hierarchy. At the same time, larger values of the inflationary scale $H_I$ in this case allow for a larger spread of the distribution $\rho(\phi)$ in the local minimum. In contrast to the CbQ regime, here we identify a large parameter region where the stochastic misalignment can naturally generate the observed DM abundance. This is in a regime where the mechanism does not require eternal inflation.
\end{itemize}

There is yet another source for a misalignment of the relaxion, which was described in~\cite{Banerjee:2018xmn}. Let us denote by $T_{b}$ the temperature, below which the barriers of the relaxion potential are established. Already at temperatures comparable to the weak scale, the Higgs vev itself changes due to the thermal corrections to the Higgs potential and, in particular, at $T\approx 160$ GeV the electroweak symmetry is expected to be restored (see e.g.~\cite{Matsedonskyi:2020mlz}). This would remove the barriers implying that $T_b$ cannot exceed that temperature, although it can be lower. If after inflation the universe reheats to temperatures $T_{\mathrm{rh}}$ well above $T_b$, the local minimum in which the relaxion is trapped will disappear for some time. Due to the slope of the potential, the relaxion will be further displaced from its minimum. Under certain conditions that were derived in~\cite{Banerjee:2018xmn}, the misalignment from this ``roll-on'' can explain the DM abundance already in the CbQ regime.

In this work, we investigate both the low-temperature and the high-temperature reheating scenarios, extending the analysis of~\cite{Banerjee:2018xmn} to the QbC regime. The largest DM window is achieved in the case of a low reheating temperature. However, even in the high reheating temperature case, we find a substantial parameter region where the displacement after reheating is small and the stochastic misalignment accounts for the DM. For completeness, we also compare these windows to the one from roll-on from~\cite{Banerjee:2018xmn} and extend the later to the QbC regime. 

Other less-minimal scenarios, in which the relaxion mechanism provides a viable DM candidate, were studied in several works. The authors of~\cite{Espinosa:2015eda} considered relaxation with two scanning scalar fields instead of one,  which enables to get away with the `coincidence problem’ of the original proposal ($\Lambda_b \lesssim $ EW scale). In this setup, the second scalar field scans the barrier $\Lambda_b$ of the relaxion potential and can be the DM of the universe. In~\cite{Banerjee:2018xmn, Banerjee:2021oeu}, a coupling of the relaxion to an additional dark photon field was added, which allows for a large DM window in the high-temperature reheating case. In~\cite{Fonseca:2018kqf}, the authors considered an alternative relaxion model where friction comes from gauge boson production. In this case, the relaxion is produced via scattering with the thermal bath and can be a warm DM candidate in the $\mathrm{keV}$ range. Let us also mention another mechanism for selecting a small Higgs mass, the sliding naturalness~\cite{TitoDAgnolo:2021nhd, TitoDAgnolo:2021pjo}, which can also explain the DM.

The outline of this work is the following. In section~\ref{ssec:abundace}, we compute the energy density stored in the coherent oscillations due to the stochastic misalignment. This allows us to construct the relaxion DM window in section~\ref{sec:dm}, for the $T_{\mathrm{rh}}<T_b$ case. We verify that in the CbQ regime the relaxion is always underabundant, as well as demonstrate how in the QbC regime the relaxion can explain the DM abundance in a large parameter region. In section~\ref{sec:highTrh}, we study the case of high reheating temperature, $T_{\mathrm{rh}} \gg T_b$, taking into account the additional displacement of the field after reheating. In section~\ref{ssec:thermal}, we estimate the thermal production of the relaxion and verify that in the DM window this contribution is negligible. In section~\ref{ssec_QCDDM}, we focus on the QCD relaxion model and the DM window in that scenario. Except for that section, we always restrict ourselves to the scenario where the relaxion mechanism does not require inflation to be eternal. We conclude in section~\ref{sec:conclusion}.

\section{Axion abundance from the stochastic misalignment}
\label{ssec:abundace}

In this section, we compute the energy density in the coherent oscillations of a generic axion-like field $\phi$ with mass $m_{\phi}$, comparing this to the observed DM abundance in the universe. We focus on the stochastic misalignment of the field i.e.~assume that the typical displacement from the local minimum of its potential is set by the inflationary Hubble scale and given by Eq.~(\ref{stochastic_variance}). We will now show, that the energy density today scales as $\Omega_{\phi,0}  \propto H_I^4 m_{\phi}^{-3/2}$. 

\bigskip

The scale of inflation in the relaxion mechanism is typically low, $H_I<v_h$, where $v_h=246$ GeV is the electroweak scale. We consider two cases, comparing the Hubble parameter at which the field starts oscillating around the minimum of its potential, $H_{\mathrm{osc}} \approx m_{\phi}/3$, to the Hubble parameter at the end of reheating,
\beq
\label{eq_HTrh}
H^2_{\mathrm{rh}} = \Bigl( \frac{1}{2t_{\mathrm{rh}}} \Bigr)^2 = \frac{8 \pi^3}{90}\frac{g(T_{\mathrm{rh}})T_{\mathrm{rh}}^4}{M_{Pl}^2},
\eeq 
where $g(T)$ denotes the effective number of relativistic degrees of freedom for the energy density~\cite{Husdal:2016haj}. If $H_{\mathrm{rh}}>H_{\mathrm{osc}}$, the field enters the oscillatory regime after reheating, in the radiation-dominated era. The condition for this can be re-written as an upper bound on the mass,
\beq
\label{H_rh_larger_H_osc}
m_{\phi} < 4 \times 10^{-5} \mathrm{eV} \Bigl( \frac{T_{\mathrm{rh}}}{100\mathrm{GeV}} \Bigr)^2 \sqrt{\frac{g(T_{\mathrm{rh}})}{100} }.
\eeq
If $H_{\mathrm{rh}}<H_{\mathrm{osc}}$, the onset of oscillations is during reheating. The relic abundance in this case is sensitive to the equation of state of the universe before reheating.

\subsection{${H_{\mathrm{rh}}>H_{\mathrm{osc}}}$}
We start with the simple case $H_{\mathrm{rh}}>H_{\mathrm{osc}}$. Employing entropy conservation, the energy density today can be expressed as,
\beq
\rho_{\phi, 0} \approx \rho_{\phi, \mathrm{osc}} \Bigl(\frac{a_{\mathrm{osc}}}{a_{0}} \Bigr)^3 \approx \frac{m_{\phi}^2 {\phi}^2}{2} \Bigl(\frac{T_{0}}{T_{\mathrm{osc}}} \Bigr)^3 \Bigl(\frac{g_{s,0}}{g_{\mathrm{s,osc}}} \Bigr).
\eeq 
Inserting $T_0 = 2.73 K$ and $T^2_{\mathrm{osc}} = M_{Pl}H_{\mathrm{osc}}(1.66\sqrt{g(T_{\mathrm{osc}})})^{-1}$ one arrives\footnote{{Here we take into account the fact that an accurate estimate for the relic density is obtained if one uses $H_{\mathrm{osc}} = m_{\phi}/A$ with $A\approx 1.6$, see e.g.~\cite{Marsh:2015xka}. This is the reason why the prefactor in Eq.~(\ref{eq:dmdensitytd}) is slightly different from the expressions in~\cite{Arias:2012az}.}} at the usual expression~\cite{Arias:2012az}
\beq
\label{eq:dmdensitytd}
\rho_{\phi, \mathrm 0}  = 6.5\, \frac{\mathrm{keV}}{\mathrm{cm}^3} \sqrt{ \frac{m_{\phi}}{\mathrm{eV} } } \Bigl( \frac{{\phi}}{10^{12}\mathrm{GeV}}\Bigr)^2  \mathcal{F}(T_{\mathrm{osc}}),
\eeq
where the dimensionless factor $\mathcal{F}(T_{\rm osc}) = ( g_{s,0} / g_{s,\mathrm{osc}}  ) (  g_{\mathrm{osc}} /g_{0}  )^{3/4}$ encodes the changing number of degrees of freedom for entropy, $g_{s}(T)$, and energy, $g(T)$~\cite{Husdal:2016haj}. It can take values between $0.3$ and $1$, thus, can be neglected for simplicity. In the case of stochastic misalignment, the typical energy density $\langle \rho_{\phi, \mathrm{osc}} \rangle = \frac{1}{2}m^2 \sigma_{\phi}^2$ from the above expression can be expressed as
\beq
\label{stochastic_1}
\frac{ \langle \Omega_{\phi, 0} \rangle  }{ \Omega_{\mathrm{DM}} } \approx  5 \sqrt{ \frac{m_{\phi}}{\mathrm{eV} } } \Bigl( \frac{\sigma_{\phi}}{10^{12}\mathrm{GeV}}\Bigr)^2  \mathcal{F}(T_{\mathrm{osc}}) \approx 20 \Bigl( \frac{\mathrm{eV} }{m_{\phi}} \Bigr)^{3/2} \Bigl( \frac{H_I}{100 \mathrm{GeV}}\Bigr)^4 \mathcal{F}(T_{\mathrm{osc}}),
\eeq
where $\Omega$ denotes the fractional energy density and $\Omega_{\mathrm{DM}}\approx0.24$ is the measured DM abundance in the universe~\cite{Aghanim:2018eyx}.

\subsection{${H_{\mathrm{rh}}<H_{\mathrm{osc}}}$}
We now move to the case $H_{\mathrm{rh}}<H_{\mathrm{osc}}$, which is slightly more complicated compared to the previous one, because here the field starts to oscillate during reheating. Some assumption should be made about the evolution of the universe at those times. For simplicity, we assume that the background energy density after inflation scales as $\rho(a) \propto H^{2}(a) \propto a^{-3(1+w)}$, where $w$ is the equation of state parameter after inflation. One may generally expect $w\approx 0$, but we will allow for a general equation of state parameter in our expressions. The energy density today can be written as
\beq
\rho_{\phi, 0} \approx \rho_{\phi, \mathrm{osc}} \Bigl(\frac{a_{\mathrm{osc}}}{a_{\mathrm{rh}}} \Bigr)^3 \Bigl(\frac{a_{\mathrm{rh}}}{a_{0}} \Bigr)^3 \approx \frac{m_{\phi}^2 {\phi}^2}{2}  \Bigl(\frac{H_{\mathrm{rh}}}{H_{\mathrm{osc}}} \Bigr)^{2/(1+w)}   \Bigl(\frac{T_{0}}{T_{\mathrm{rh}}} \Bigr)^3 \Bigl(\frac{g_{s,0}}{g_{\mathrm{s,rh}}} \Bigr).
\eeq 
Performing the same steps as in the previous case we arrive at
\beq
\langle \rho_{\phi, 0} \rangle \approx 6.5\, \frac{\mathrm{keV}}{\mathrm{cm}^3} \sqrt{ \frac{m_{\phi}}{\mathrm{eV} } } \Bigl( \frac{\sigma_{\phi}}{10^{12}\mathrm{GeV}}\Bigr)^2   \Bigl( \frac{H_{\mathrm{rh}}}{H_{\mathrm{osc}}} \Bigr)^{\frac{1-3w}{2(1+w)}} = \langle \rho^{(w=1/3)}_{\phi, 0} \rangle  \Bigl( \frac{H_{\mathrm{rh}}}{H_{\mathrm{osc}}} \Bigr)^{\frac{1-3w}{2(1+w)}}.
\eeq
Here $\rho^{(w=1/3)}_{\phi, 0}$ can be understood as today's energy density in the case of $w=1/3$ which is also the prediction for the relic density in the previous case of $H_{\mathrm{rh}}>H_{\mathrm{osc}}$. As can be seen, $w<1/3$ leads to a suppression of the relic density. 

\bigskip

Combining the two cases 
${H_{\mathrm{rh}}<H_{\mathrm{osc}}}$
and ${H_{\mathrm{rh}}>H_{\mathrm{osc}}}$
the typical DM fraction can be expressed as
\beq
\label{relaxion_dm_abund}
\boxed{ \frac{ \langle  \Omega_{\phi, 0} \rangle }{ \Omega_{\mathrm{DM}} }  \approx   20\,  \Bigl( \frac{\mathrm{eV} }{m_{\phi}} \Bigr)^{3/2} \Bigl( \frac{H_I}{100 \mathrm{GeV}}\Bigr)^4  \: \mathrm{min} \Bigl\{ 1,  \Bigl( \frac{H_{\mathrm{rh}}}{H_{\mathrm{osc}}} \Bigr)\Bigr\}^{\frac{1-3w}{2(1+w)}}  .}
\eeq
For fixed values of $w$ and $T_{\mathrm{rh}}$, $\langle  \Omega_{\phi, 0} \rangle/\Omega_{\mathrm{DM}}$ is determined by the  the mass $m_{\phi}$ and has a strong dependence on the Hubble scale during inflation $H_I$.

\section{Relaxion dark matter for $T_{\mathrm{rh}} \lesssim T_b$}
\label{sec:dm}

\begin{figure}[!t]
	\centering
	\includegraphics[width=0.9\textwidth]{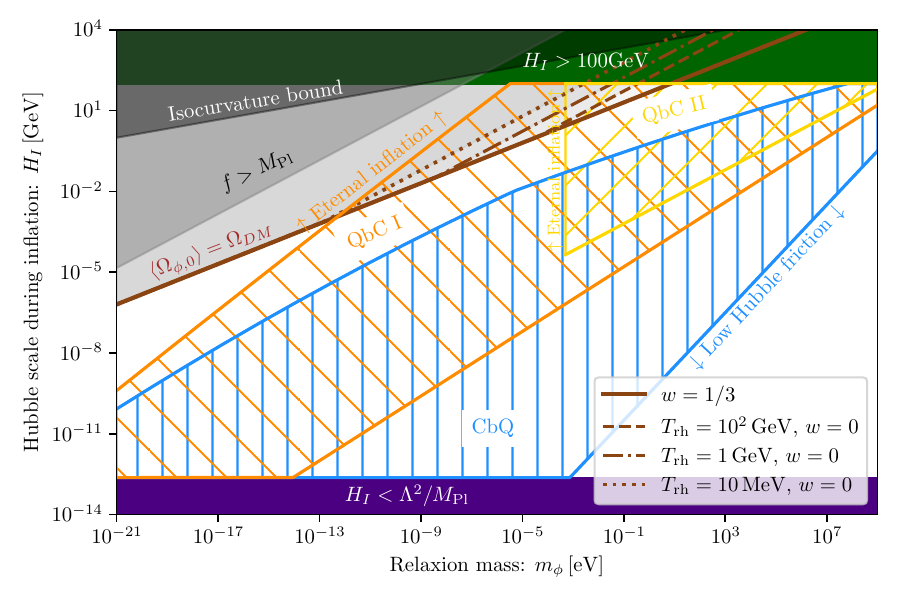}
	\caption{The relaxion parameter region in the $[m_{\phi}, H_I]$ plane. The brown lines show where the stochastic misalignment of the relaxion would explain DM according to Eq.~(\ref{relaxion_dm_abund}). The different lines correspond to different reheating temperatures and values of the equation of state parameter between the end of inflation and the end of reheating ($w=1/3$ for the solid line and $w=0$ for the rest). The region hashed with blue vertical lines corresponds to the CbQ regime and does not overlap with these brown lines. In contrast, the relaxion in the QbC I (the region hashed with orange lines) and QbC II (the region hashed with yellow lines) regimes has such an overlap, which is where it can naturally explain DM.}
	\label{DMHm}
\end{figure}

The stochastic misalignment of the relaxion can naturally explain the observed DM abundance in a large parameter region if the QbC regime is included. This is illustrated in Fig.~\ref{DMHm}, where we show the allowed parameter region for the relaxion and the brown lines are determined from Eq.~(\ref{relaxion_dm_abund}) setting the DM fraction to one. We consider several reheating temperatures for the $w=0$ equation of state before reheating, as well as the $w=1/3$ case where the relic density does not depend on $T_{\mathrm{rh}}$. 
In section~\ref{ssec_insufDM_CbQ} we explain why the relaxion is always underabundant in the CbQ regime. We then construct the DM window in section~\ref{ssec_DM_window}.
Here we focus on the $T_{\mathrm{rh}} \lesssim T_{b}$ scenario, for which the misalignment of the relaxion is unaffected by the reheating of the universe. 
The scenario $T_{\mathrm{rh}} \gg T_{b}$ is examined in section \ref{sec:highTrh}.

Before proceeding, let us summarize the constraints that are imposed on the relaxion when constructing its parameter region. We refer to~\cite{Chatrchyan:2022pcb} for more details. The main free parameters are $g$ (we set $g'=g$), $\Lambda$, $f$ and $H_I$. 
The relaxion is expected not to back-react on Hubble expansion during inflation, implying $H_I^2>(8\pi/3)\Lambda^4/M_{\mathrm{Pl}}^2$. 
The spread in the Higgs mass due to diffusion effects $\Delta \mu_h^2\sim H_I^2$ is supposed to be small and we impose $H_I < 100 \mathrm{GeV}$. 
The separation between the local minima is also required to be less than the scanning precision, $g'\Lambda (2\pi f) <|\mu_h^2| = (88\mathrm{GeV})^2$. 
The decay constant is usually assumed to be subplanckian, $f < M_{\mathrm{Pl}}$, and we also require $f>\Lambda$ for the consistency of the effective theory.
The cut-off scale $\Lambda$  is assumed to be at least $\mathrm{TeV}$.
Except for section~\ref{ssec_QCDDM}, we always require that inflation is not eternal i.e.~that the minimal number of e-folds required for the field to relax the cut-off scale, $N_I \sim 3H_I^2/(g^2\Lambda^2)$, does not exceed the critical number of e-folds corresponding to eternal inflation $N_c = (2\pi^2/3)M_{\mathrm{Pl}}^2/H_I^2$~\cite{Graham:2018jyp, Dubovsky:2011uy}. This condition translates into
\beq
\label{noneternal}
g\Lambda> \frac{3H_I^2}{\sqrt{2}\pi M_{Pl}}.
\eeq 
We stress that in the CbQ regime this condition is satisfied automatically.

The parameter $\Lambda_b$, by which we denote the barrier height at the final local minimum corresponding to the correct Higgs vev, is determined from the stopping condition. As it was derived in~\cite{Chatrchyan:2022pcb}, inflationary fluctuations allow for transitions of $\phi$ between neighbouring local minima and, in the general case, the relaxion stops at a minimum for which the Hawking-Moss transition rate to the next minimum is suppressed. We use
\beq
\label{stopping_condition}
B = \frac{8\pi^2\Delta V_b^{\rightarrow}}{3H_I^4} \sim 1
\eeq
as the stopping condition, where $\Delta V_b^{\rightarrow}$ denotes the height of the barrier to the next minimum. We note that
\begin{itemize}
	\item In the CbQ regime, this condition always implies $\Lambda_b^4 \approx g\Lambda^3f$. Here we do not consider the scenario of low Hubble friction by requiring $H_I>\dot \phi_{\mathrm{SR}}/(2\pi f)$. This ensures that the relaxion tracks the slow-roll velocity $\dot \phi_{\mathrm{SR}} = g\Lambda^3/3H_I$ as it rolls down the potential. If this is not the case, the field can stop in a much deeper minimum as well as undergo fragmentation, as it was explained in~\cite{Fonseca:2019lmc}.
	
	\item In the QbC regime we have separated a QbC I regime where the relation $\Lambda_b^4 \approx g\Lambda^3f$ still holds, and a QbC II regime in which Eq.~(\ref{stopping_condition}) implies a barrier height that is determined by the inflationary Hubble scale, $\Lambda_b^4 \approx (3/16\pi^2)H_I^4$ or $\Lambda_b \sim H_I$. The transition between the QbC I and the QbC II regimes occurs approximately at $H_I^4 = (16\pi^2/3)g\Lambda^3f$. 
\end{itemize}

In principle, the stopping condition depends on the total number of e-folds $N_I$ of inflation, i.e.~the larger $N_I$ the deeper would be the final minimum of the relaxion. This dependence is however only logarithmic and hence can be neglected. 
An upper bound 
\begin{equation}
\Lambda_b < \sqrt{4\pi}v_h,
\end{equation}
is imposed to ensure that the barrier height is sensitive to the Higgs vev.

Finally, it is convenient to introduce the dimensionless parameter $\delta$, defined as
\beq
\cos\delta = \frac{g\Lambda^3f}{\Lambda_b^4}.
\eeq
As explained in~\cite{Chatrchyan:2022pcb}, the mass of the relaxion, the barrier height and the separation between the minimum $\phi_0$ and the maximum of the barrier $\phi_b$ are given respectively by
\beq
m_{\phi}^2 = \frac{\Lambda_b^4}{f^2} \times \sin\delta, \:  \:  \:  \:  \:   \:  \:  \:  \:  \Delta V_b^{\rightarrow} = 2\Lambda_b^4\times[\sin\delta - \delta \cos\delta],  \:  \:  \:  \:  \:   \:  \:  \:  \: \phi_b - \phi_0 = 2f \times \delta,
\eeq
where the $\delta$-dependent corrections are due to the linear slope. Near the first local minimum $\delta \ll 1$ and, as a consequence, both the mass and the barrier height are suppressed compared to the naive expectation (see also~\cite{Banerjee:2018xmn}).

\subsection{Insufficient dark matter in the CbQ regime}
\label{ssec_insufDM_CbQ}

We start demonstrating that the relic density from the stochastic relaxion misalignment in the CbQ regime is too small to explain DM. 

\bigskip

In the CbQ regime, $H_I<g^{1/3}\Lambda$ holds and the relaxion stops near $\Lambda_b^4 \approx g\Lambda^3 f$. Assuming that $w\leq 1/3$ holds during reheating, one can write for the relic density
\beq
\frac{ \Omega_{\phi, 0} }{ \Omega_{\mathrm{DM}} }  \leq 20  \Bigl( \frac{m_{\phi}}{\mathrm{eV} } \Bigr)^{-3/2} \Bigl( \frac{H_I}{100 \mathrm{GeV}}\Bigr)^4 < 20 \Bigl( \frac{m_{\phi}}{\mathrm{eV} } \Bigr)^{-3/2} \Bigl( \frac{\Lambda_b^{4/3}f^{-1/3}}{100 \mathrm{GeV}}\Bigr)^4.
\eeq
Using $\sin\delta \geq  \Lambda_b^2/(\Lambda \sqrt{-\mu_h^2})$ and inserting the maximal value for the cut-off scale~\cite{Chatrchyan:2022pcb} one arrives at
\beq
\frac{ \Omega_{\phi, 0} }{ \Omega_{\mathrm{DM}} } < 8.3  \times 10^{-8} \Bigl( \frac{m_{\phi}}{\mathrm{eV} } \Bigr)^{-1/6} \Bigl( \frac{\Lambda_b}{\sqrt{4\pi} v_h} \Bigr)^{4/3} \Bigl( \frac{\Lambda}{4\times 10^{9}\mathrm{GeV}} \Bigr)^{2/3},
\eeq
and, even for the lightest possible masses of fuzzy DM $m_{\phi} \sim 10^{-21} \rm eV$, the relic density does not exceed $\Omega_{\phi, 0}\sim 10^{-4}$, even if we allow for super-Planckian decay constants. Note that at those light masses an even stronger bound on the relic density can be put from the $\Lambda<f$ condition. To illustrate this, in figure~\ref{DMHm} we show the relaxion parameter region in $H_I$ vs $m_{\phi}$ plane. The CbQ region, shown in blue, indeed does not overlap with the
$\Omega_{\phi, 0} ={ \Omega_{\mathrm{DM}} }$ lines, in contrast to the QbC I (orange) and QbC II (yellow) regions that we discuss next.

\subsection{Dark matter from stochastic misalignment (QbC)}
\label{ssec_DM_window}

To obtain the bounds on the relaxion mass in the DM window we set $\langle \Omega_{\phi, 0} \rangle$ equal to $\Omega_{\mathrm{DM}}$ in Eq.~(\ref{relaxion_dm_abund}). Assuming $w = 1/3$ one arrives at
\beq
m_{\phi}  \approx 10\, \mathrm{eV} \Bigl( \frac{H_I}{100\mathrm{GeV}}\Bigr)^{8/3}.
\eeq
If one instead uses $w=0$, the expression for the mass takes the form
\beq
m_{\phi} \approx 0.4 \, \mathrm{eV} \Bigl( \frac{H_I}{100 \mathrm{GeV}} \Bigr)^{2}  \Bigl( \frac{T_{\mathrm{rh}}}{100\mathrm{GeV}} \Bigr)^{1/2} \Bigl( \frac{g(T_{\mathrm{rh}})}{100} \Bigr)^{1/8} .
\eeq
For the upper bound we simply impose $H_I<100$ GeV for the inflationary Hubble scale in the above expressions. Note that for the considered reheating temperatures, the upper bound depends on physics before reheating, which is consistent with the fact that the onset of oscillations for masses that do not satisfy (\ref{H_rh_larger_H_osc}) is before reheating.

A lower bound on the mass of relaxion DM can be imposed by requiring (\ref{noneternal}) to avoid eternal inflation. Here the stopping condition near the first minimum is relevant and the $\delta$-dependent prefactor in the expression for the mass, which is now expected to be small, should be included. Inserting everything into the expression for the mass one can write
\beq
m_{\phi}^2 = \frac{\Lambda_b^4}{f^2}\sin\delta \approx \frac{\Lambda_b^6}{f^2\Lambda (-\mu_h^2)^{1/2}} > \Bigl( \frac{3H_I^2}{\sqrt{2}\pi M_{Pl}}\Bigr)^{3/2}\frac{\Lambda^2}{f^{1/2} (-\mu_h^2)^{1/2}}.
\eeq
Rewriting this as an upper bound on $H_I$ and inserting into the expression for the relic density with $\Omega_{\phi, 0} \sim \Omega_{\mathrm{DM}}$ one arrives at
\beq
m_{\phi}> 10^{-13} \mathrm{eV} \Bigl( \frac{\Lambda}{\mathrm{TeV}} \Bigr)^{\frac{16}{7}}\Bigl( \frac{M_{Pl}}{f} \Bigr)^{\frac{4}{7}}.
\eeq
For this lower bound the oscillations start after reheating, according to (\ref{H_rh_larger_H_osc}).

\bigskip

To summarize, in a wide range of masses,
\beq
\label{mass_range_DM}
\boxed{ 
	10^{-13}  \Bigl( \frac{\Lambda}{\mathrm{TeV}} \Bigr)^{\frac{16}{7}}\Bigl( \frac{M_{Pl}}{f} \Bigr)^{\frac{4}{7}} < \frac{m_{\phi}}{\mathrm{eV}} < 0.4 \times  10^{4w} \Bigl( \frac{H_I}{100 \mathrm{GeV}} \Bigr)^{2(1+w)}  \Bigl[ \frac{T_{\mathrm{rh}}}{100\mathrm{GeV}}  \Bigl( \frac{g(T_{\mathrm{rh}})}{100} \Bigr)^{\frac{1}{4}} \Bigr]^{\frac{1-3w}{2}}  ,} 
\eeq
the relaxion can potentially constitute the DM. It is worth mentioning that the DM window includes not only the regime of late stopping, QbC II, but also the regime where the relaxion stops closer to the first minimum, QbC I. In the second case, the typically small misalignment from the minimum is compensated by the large value of $f$ (similar to the stochastic QCD axion scenario from~\cite{Graham:2018jyp}). We note that the regions above/below the $\langle \Omega_{\phi, 0} \rangle = \Omega_{\mathrm{DM}}$ are not strictly excluded. The relaxion would simply need by chance to sit very close to the minimum/maximum of the potential to generate the required abundance.

\bigskip

\begin{figure}[!t]
	\centering
	\includegraphics[width=0.8\textwidth]{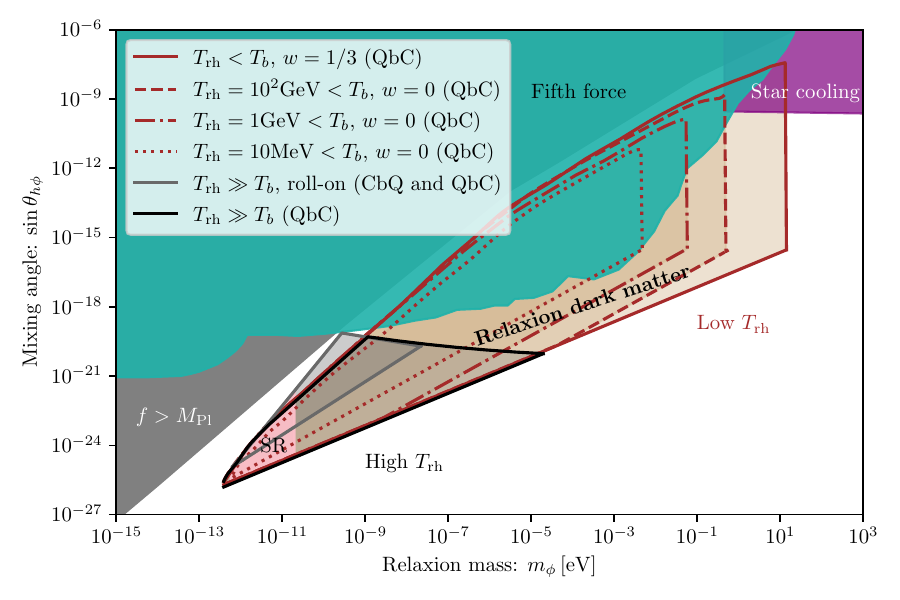}
	\includegraphics[width=0.8\textwidth]{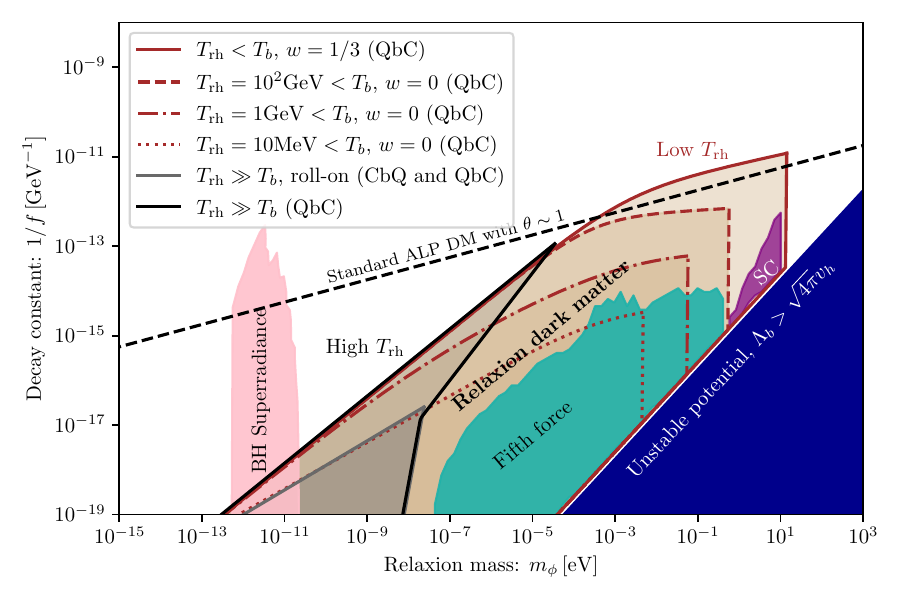}
	\caption{The relaxion DM window in the $[m_{\phi}, \sin\theta_{h\phi}]$ (top) and $[m_{\phi}, 1/f]$ (bottom) planes. The brown shaded regions correspond to the stochastic window in the QbC regime with $T_{\mathrm{rh}}<T_b$. Here different lines correspond to different values of the equations of state parameter during reheating and different values of the reheating temperature. The grey region shows the DM window from roll-on for $T_{\mathrm{rh}} \gg T_b$, which was proposed in~\cite{Banerjee:2018xmn} for the CbQ regime, and extended here for the QbC case. The stochastic window in the QbC regime for $T_{\mathrm{rh}} \gg T_b$ is enclosed by the black solid line. The constraints from fifth force experiments~\cite{Flacke:2016szy} (navy), stellar cooling~\cite{Hardy:2016kme} (purple) as well as from black hole superradiance~\cite{Baryakhtar:2020gao} (pink) are shown for the DM window.}
	\label{DM1}
\end{figure}

\begin{figure}[!t]
\centering
\includegraphics[width=0.8\textwidth]{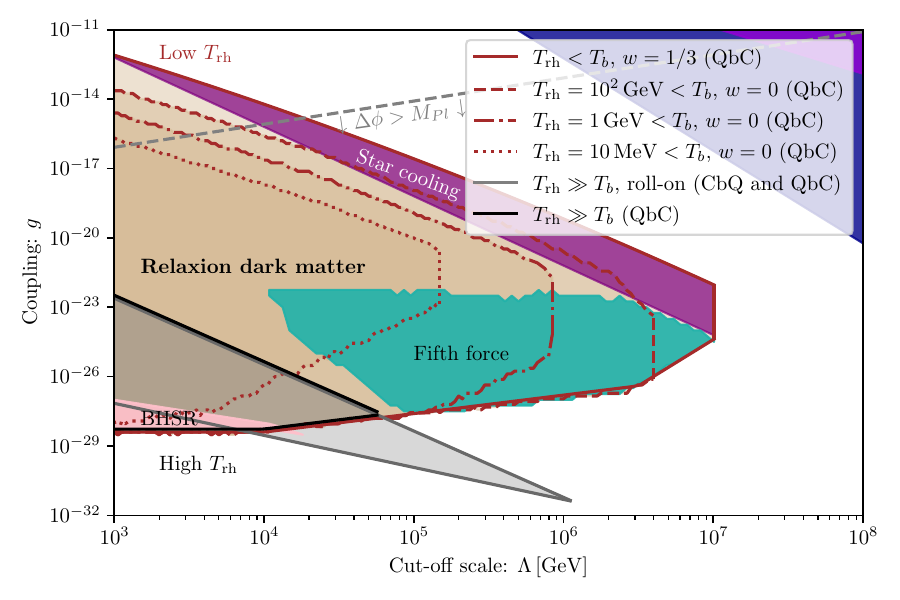}
\includegraphics[width=0.8\textwidth]{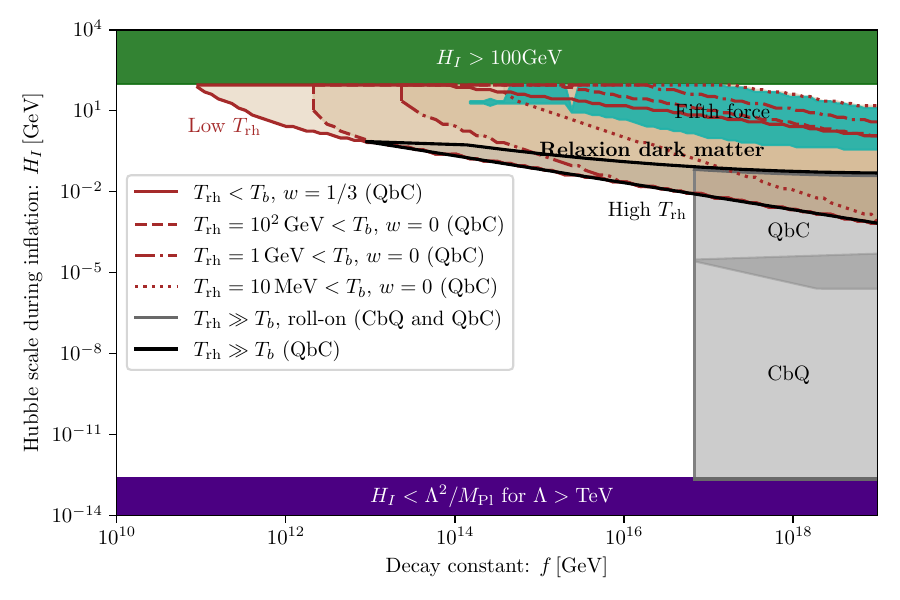}
\caption{The relaxion DM windows (in brown and grey), in the $[\Lambda, g]$ (top) and the $[f, H_I]$ (bottom) planes, complementing figure~\ref{DM1}.}
\label{DM2}
\end{figure}

We plot the parameter space where it is possible for the relaxion to generate the correct DM abundance in Figs.~\ref{DM1} and~\ref{DM2}, in four different planes. In figure~\ref{DM1} the parameter region is shown in the $\sin\theta_{h\phi}$ vs $m_{\phi}$ plane (upper panel), as well as in the $f^{-1}$ vs $m_{\phi}$ plane (lower panel). Here, the following definition of the mixing angle is used (see also~\cite{Banerjee:2020kww, Chatrchyan:2022pcb}),
\beq
\sin\theta_{h\phi} \approx  - \frac{1}{m_h^2} \frac{\partial^2 V}{\partial h \partial \phi}\Big|_{v_h, \phi_{\mathrm{0}}},
\eeq
where $V(h, \phi)$ is given in Eq.~(\ref{eq_potential}). The DM window is highlighted in brown. We use the same choices of $w$ and $T_{\mathrm{rh}}$ as in figure~\ref{DMHm}. Constraints arising from fifth force experiments, including inverse-square-law and equivalence-principle tests~\cite{Hoskins:1985tn, Kapner:2006si, Schlamminger:2007ht, Bordag:2009zz, Chen:2014oda, Berge:2017ovy, PhysRevLett.124.051301}, are shown in navy, while the stellar cooling bounds due to resonant production in the plasma \cite{Hardy:2016kme} are shown in purple. The limits from~\cite{Dev:2020jkh,Balaji:2022noj}, which consider the bremsstrahlung production, lead to much stronger bounds on the mixing angle. However, those have been questioned by Caputo et al in
\cite{Caputo2023}. Both the fifth force constraints and the stellar bounds are based on the mixing angle with the Higgs. They are also projected in the remaining plots of both figures. In other words, a point inside the DM window is marked as excluded, if it is excluded for any choice of the remaining free parameters, including the choice of the reheating temperature for $w=0$, that can explain DM. Similarly, we include the black hole superradiance bounds from~\cite{Baryakhtar:2020gao}, which exclude the $10^{-12}\,\mathrm{eV} \lesssim m_{\phi}\lesssim 10^{-11}\, \mathrm{eV}$ masses for the relaxion DM.\footnote{ALP DM with masses in the range $10^{-15}\mathrm{eV}\lesssim m_a \lesssim 10^{-13}\mathrm{eV}$ may leave unique signatures in direct detection experiments~\cite{Kim:2021yyo}. In the case of relaxion DM such a mass range requires either $f>M_{\mathrm{Pl}}$ or eternal inflation.} The dashed line in the $f^{-1}$ vs $m$ plane of Fig.~\ref{DM1} corresponds to the standard ALP DM window with $\theta \sim 1$, which can be computed using Eq.~(\ref{eq:dmdensitytd}), setting $\phi^2 \sim f^2$.

In Fig.~\ref{DM2}, the parameter region is shown in the $g$ vs $\Lambda$ (upper panel) and the $H_I$ vs $f$ (lower panel) planes. As can be inferred from the upper panel, the DM relaxion can explain cut-off scales as large as almost $10^{7}$ GeV. However, taking into account the constraints from fifth force experiments and from astrophysics, the cut-off can reach up to around $10^5\mathrm{GeV}$. Note that the decay constant for such relaxion can be as small as $10^{11}-10^{14}\, \mathrm{GeV}$ and the Hubble scale during inflation at most $100$ GeV. We  also find that $\Lambda_b$ can take values in the range $1\mbox{ GeV}<\Lambda_b<\sqrt{4\pi} v_h$ in the DM window. Below the dashed line in the $g$ vs $\Lambda$ plane the typical relaxion excursion $\Delta \phi$ during inflation exceeds the Planck scale. As can be seen, such superplanckian field excursion can be avoided in the upper region of the parameter space for our relaxion DM window.

In the upper right region of the $H_I$ vs $f$ plane in figure~\ref{DM2}, above the DM window, the relaxion overproduces DM, hence, this region is excluded. In contrast, there are no regions in the remaining three panels of figures~\ref{DM1} and~\ref{DM2} that are excluded by overproduction for any choice of the remaining free parameters.

\bigskip

\textbf{Bounds on isocurvature fluctuations:} The relaxion picks up the Hubble-sized isocurvature fluctuations, $\delta \phi \sim H_I$, during inflation. These perturbations are uncorrelated with the adiabatic ones and modify the temperature power spectrum of the cosmic microwave background (CMB). So far, they have not been observed by Planck, and, therefore, are tightly constrained. We use the bound from~\cite{DiLuzio:2020wdo},
\beq
\label{iso}
\frac{H_I}{\mathrm{GeV}} < 0.3 \times 10^{7} \frac{\phi}{10^{11} \mathrm{GeV}} \Bigl( \frac{\Omega_{DM}}{\Omega_{\phi, 0}} \Bigr) ,
\eeq
which approximates the potential as quadratic. The last assumption is justified in our case since the relaxion can get trapped only in a long-lived minimum, which necessarily has $\sigma_{\phi} < f$ such that the anharmonicities of the potential are not so important. 

The isocurvature bound, computed using Eq.~(\ref{iso}) is shown in Fig.~\ref{DMHm}. As can be seen, it is weaker compared to the bound from overproduction.

\section{Relaxion dark matter for $T_{\mathrm{rh}}\gg T_b$}
\label{sec:highTrh}

We now move to the case of high reheating temperatures, $T_{\mathrm{rh}}\gg T_b$. The barriers temporarily disappear in this case and the relaxion can roll down further along its potential during that time interval. Here one has to make sure that the field gets trapped again once the barriers are back. The displacement was computed in~\cite{Banerjee:2018xmn}, and in section~\ref{ssec_displacement} we revisit the computation generalizing it to the QbC regime. As it was seen in the previous section, the stochastic misalignment of the relaxion cannot explain the observed DM abundance in the universe in the CbQ regime. This is however not true when $T_{\mathrm{rh}}\gg T_b$. In the later case, the additional displacement of the relaxion can itself generate the required misalignment to explain DM as it was found in~\cite{Banerjee:2018xmn}. We discuss this DM window in section~\ref{ssec_DMroll}, first in the CbQ regime and then extend it to the QbC regime. Finally, in section~\ref{ssec_DMstochastic2}, we construct the stochastic DM window for the case of high reheating temperatures, where we require that the displacement during reheating is much smaller compared to the stochastic misalignment.

\subsection{The displacement (roll-on) after inflation}
\label{ssec_displacement}

\subsubsection*{CbQ and QbC I}
The displacement of the relaxion after its barriers disappear was studied in~\cite{Banerjee:2018xmn} for the CbQ regime. The field evolution can be studied by solving the field equations of motion in the radiation-dominated universe,
\beq
\label{eom}
\ddot \phi + \frac{3}{2t}\dot \phi -g\Lambda^3 + C(T) \frac{\Lambda_b^4}{f} \sin\Bigl(\frac{\phi}{f}\Bigr)= 0.
\eeq
The function $C(T)$ encodes the temperature-dependence of the barriers of the potential. For simplicity here it is taken to be a step function, $C(T(t)) = \theta(T_{b}/T(t) - 1)$, implying that the barriers reappear instantaneously at $T_{b}$. 

We first discuss the displacement at $t<t_b$, assuming that the relaxion starts at rest in its local minimum $\phi_0$ at $t= t_{\mathrm{rh}}$. For $C=0$ the solution to the above equation is given by $\dot \phi(t) = \frac{2}{5} g\Lambda^3 t [1- (t_{\mathrm{rh}}/t)^{5/2}]$. If $T_{\mathrm{rh}} \gg T_b$, one can approximate $\dot \phi(t_b) \approx \frac{2}{5} g\Lambda^3 t_b $ and, therefore, $\phi(t_b)  - \phi_0 \approx  \frac{1}{5} g\Lambda^3 t_b^2 $. It is convenient to introduce a dimensionless parameter
\beq
\R = \frac{g\Lambda^3t_b^2}{f},
\eeq
which depends on the slope $g\Lambda^3$, the decay constant and the time $t_b$ at which the barriers reappear. It characterizes the typical displacement in units of $f$ during the time when the barriers are absent, $\Delta \phi = \R f/5$.

Next, we consider the evolution at $t>t_b$, after the barriers reappear. The average slope of the potential from the minimum to the maximum can be estimated as
\beq
\frac{\delta V}{\delta \phi} \approx \frac{\Delta V_b^{\rightarrow}}{\phi_b - \phi_0} = \frac{2\Lambda_b^4 [\sin\delta - \delta \cos\delta]}{2\delta f} = g\Lambda^3\Bigl[\frac{\tan \delta}{\delta} - 1\Bigr].
\eeq
For $\delta \ll 1$, which always holds in the CbQ regime, the sum of the last two terms in Eq.~(\ref{eom}) is then approximately $\delta V/\delta \phi \approx g\Lambda^3 \delta^2/3$, which is much smaller compared to each of the first two terms in Eq.~(\ref{eom}) at $t=t_b$. The solution to the equation of motion ignoring the potential terms has the form $\phi(t_b)  - \phi_0 = g\Lambda^3t_b^2[1-\frac{4}{5}\sqrt{t_b/t}]$ or, equivalently, $\dot \phi \propto a^{-3}$ for $t>t_b$.

Combining everything, the total displacement can be expressed as
\beq
\label{Hyungjin_theta}
\Delta \phi \approx \R f= \frac{ g\Lambda^3 }{ 4H_b^2 } \: \: \: \: \: \: \: \: \: \: \: \: (\text{total displacement, }\: \delta \ll 1).
\eeq

In order to get trapped, the displacement of the relaxion has to be less than the distance to the next maximum,
\beq
\label{Hyungjin_get_trapped}
\Delta \phi < 2\delta f.
\eeq
If this is not the case, the relaxion would runaway since the acceleration of the relaxion in the regions after a maximum and before its next minimum will not be compensated by the deceleration in the much narrower regions from a minimum to its next maximum.

The onset of oscillations is at $H_{\mathrm{osc}} = m_{\phi}/3$, where $m_{\phi}^2 = ({ \Lambda_b^4}/{f^2}) \sin \delta = ({ g\Lambda^3}/{f}) \tan \delta$. In the case when the relaxion gets re-trapped, the onset of oscillations is much later compared to the reappearance of the barriers at $H_b$, which follows directly from Eq.~(\ref{Hyungjin_get_trapped}) for $\delta \ll 1$. The field thus remains frozen at the displacement determined by Eq.~(\ref{Hyungjin_theta}) until eventually it starts oscillating around the minimum.
\bigskip

The above discussion of the dynamics at $t>t_{b}$ assumes the relation $\Lambda_b^4 \approx g\Lambda^3 f$ or, equivalently, $\delta \ll 1$ for the local minimum of the relaxion. This condition is however not satisfied in the QbC II regime. We now generalize the discussion to this case with $\tan \delta \gg 1$, splitting it into several parts. 

\subsubsection*{QbC II}

\begin{itemize}
\item $\R < 9/(4\tan\delta)$: In this case, the displacement of the relaxion at $t_b$ is much smaller compared to the distance to the next maximum. Within the harmonic approximation the equation of motion can be written as $\ddot \phi + \frac{3}{2t}\dot \phi  + m_{\phi}^2  \phi= 0$.  In this regime $H_{\mathrm{osc}}<H_b$ holds and the mass term in the equation is small (at least at $t_b$) compared to the other two terms. The solution is hence similar to the one from~\cite{Banerjee:2018xmn} with $\dot \phi \propto a^{-3}$ and the total displacement $\Delta \phi$ approximately given by Eq.~(\ref{Hyungjin_theta}). The relaxion gets trapped in the same minimum.

\item $9/(4\tan\delta) < \R < 10 \delta$: In this case, $H_{\mathrm{osc}} < H_b$, therefore, the onset of oscillations is directly once the barriers are back at $t_b$. The friction term is subdominant already at $t_b$ and the oscillation amplitude decreases as $a^{-3/2}$. The kinetic energy at $t_b$ is suppressed compared to potential energy. Consequently, the maximal displacement can be estimated as
\beq
\label{theta_2}
\Delta \phi \approx \frac{\R f}{5} = \frac{ g\Lambda^3 }{ 20H_b^2 } \: \: \: \: \: \: \: \: \: \: \: \: (\text{total displacement, }\: \delta \gg 1).
\eeq
The upper bound on $ \R$ ensures that $\Delta \phi < 2\delta f$ and the relaxion again does not overshoot to the next local minimum.

\item $10 \delta < \R < 25/\cos\delta $: Here the relaxion finds itself in a local minimum different from the original one at $t_b$. Importantly, it is not guarantied that the relaxion gets trapped in that minimum afterwards. Whether this happens or not depends on the value of $\R$ i.e.~whether the energy density at $t_b$ exceeds the barrier height. We generally expect an $\mathcal{O}(1)$ misalignment from the local minimum with the oscillations starting directly once the barriers are back.

\item $\R > 25/\cos\delta $: In this case, the kinetic energy of the relaxion is always enough to overcome the next barrier, even if it stops exactly at a local minimum at $t_b$. As a result, the relaxion keeps overshooting and rolling down.
\end{itemize}

The requirement for the relaxion to get re-trapped puts additional constraints on the parameter space compared to the low-temperature reheating scenario. In particular, it entirely excludes the QbC II regime. Indeed, the condition $\R<25/\cos\delta$ to get re-trapped can be re-formulated as a lower bound on $H_{b}$ and, requiring non-eternal inflation according to Eq.~(\ref{noneternal}) and using the stopping condition of QbC II, one arrives at
\beq
H_b > \frac{g\Lambda^3}{10\Lambda_b^2} > 2\sqrt{6} \frac{\Lambda^2}{M_{\mathrm{Pl}}}.
\eeq
Even for a cut-off scale of $\Lambda = \mathrm{TeV}$ this exceeds the Hubble scale corresponding to a temperature $T_b = 100$  GeV. It is therefore impossible for the relaxion to get re-trapped in this regime. 

In the CbQ and QbC I regimes the condition to get trapped reduces the available parameter region as compared to the $T_{\mathrm{rh}}<T_b$ case. The condition $\R<2\delta$ can be expressed as an upper bound on $g$. Setting $f=M_{\mathrm{Pl}}$ and assuming that the relaxion stops at the first minimum, which implies $\delta = {\Lambda_b^2}/({\Lambda \sqrt{-\mu_h^2}})$, one arrives at
\beq
\label{eq:destabilization}
g < 64 (8\pi^3)^2 \frac{T_b^8}{\Lambda^5 M_{\mathrm{Pl}}\mu_h^2} \: \: \: \: \: \: \: \: \: \: \text{and} \: \: \: \: \: \: \: \: \: \: g < 8 (8\pi^3) \frac{T_b^6}{\Lambda^4 M_{\mathrm{Pl}}\mu_h}.
\eeq
where $\Lambda_b^4 \approx g\Lambda^3f$ and $\Lambda_b<T_b$ was used, respectively.

\begin{figure}[!t]
\centering
\includegraphics[height=0.308\textheight]{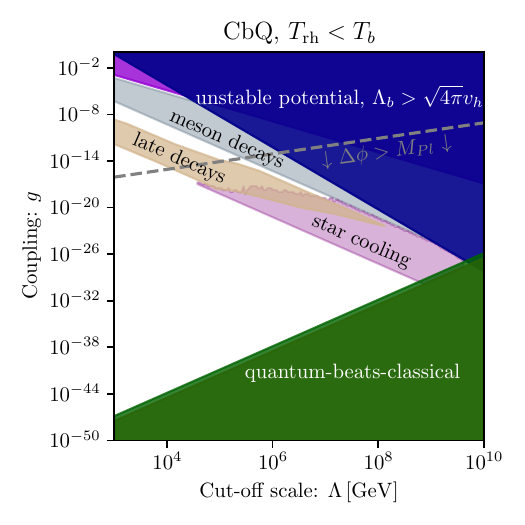}
\includegraphics[height=0.308\textheight]{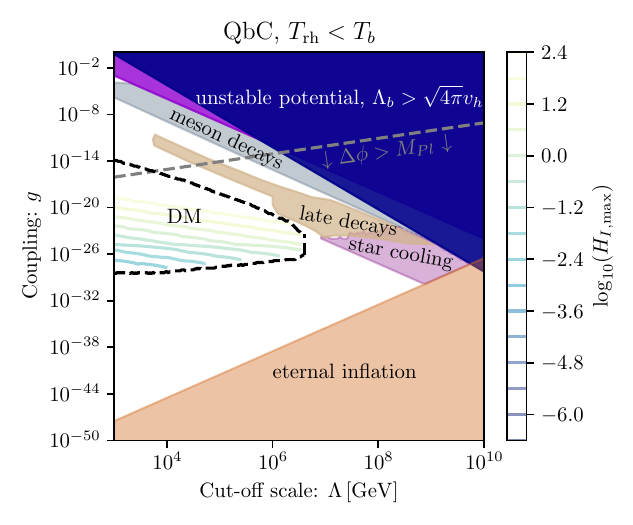}
\includegraphics[height=0.308\textheight]{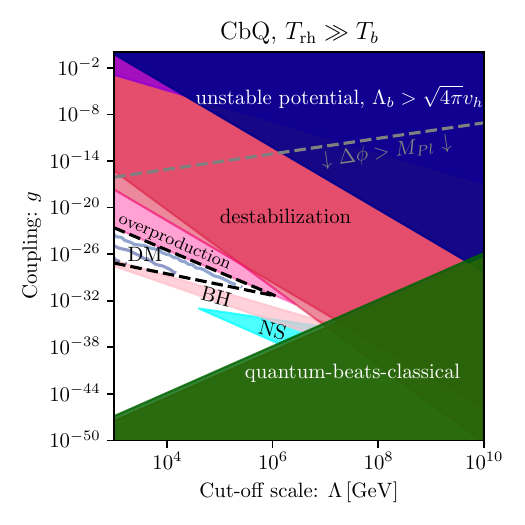}
\includegraphics[height=0.308\textheight]{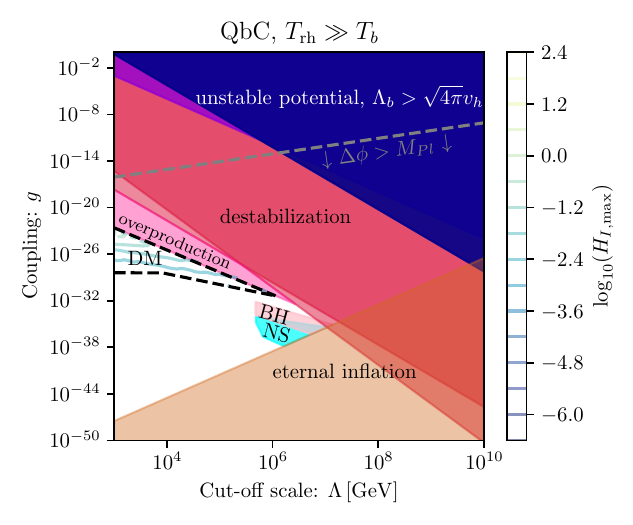}
\caption{The relaxion parameter region in the $[\Lambda, g]$ plane for the CbQ (left) and QbC (right) regimes for small (top) and large (bottom) reheating temperatures. Constraints from meson decays, stellar cooling,
 late decays ($1s < \tau_{\phi} < 10^{26}s$), black hole superradiance and density-induced runaway (in NSs) are incorporated. The region where the relaxion can explain DM is inside the black dashed lines, where also the contours of $\log_{10}(H_{I, \mathrm{max}})$ are shown. In the low-temperature reheating scenario $w=0$ before reheating is assumed. The laboratory and the astrophysical constraints under the additional assumption that the relaxion explains DM are not shown here and can be found in the upper panel of Fig.~\ref{DM2}.}
\label{fig:summary}
\end{figure}

The viable parameter region for the relaxion is shown in white in Fig~\ref{fig:summary} in the $[\Lambda, g]$ plane. The CbQ (left) and QbC (right) cases are separated, as well as the low-temperature (top) and the high-temperature (botttom) reheating scenarios.  In the  $T_{\mathrm{rh}}\gg T_{b}$ case, the red region, marked ``destabilization'', is excluded by the constraints from Eq.~(\ref{eq:destabilization}) for any $T_{b} < 100$ {GeV}. Shaded are also the regions excluded by proton beam dump and accelerator experiments testing meson decays~\cite{BNL-E949:2009dza, Belle:2009zue,BaBar:2013npw,LHCb:2015nkv, LHCb:2016awg, CHARM:1985anb} (grey), stellar cooling bounds~\cite{Hardy:2016kme} (purple), cosmological bounds on late relaxion decays via the Higgs mixing for lifetimes of $1s < \tau_{\phi} < 10^{26}s$ ~\cite{Flacke:2016szy}, black hole superradiance~\cite{Baryakhtar:2020gao} (pink) and runaway in stars induced by finite-density effects~\cite{Balkin:2021wea} (cyan). More details about the various constraints can be found in~\cite{Chatrchyan:2022pcb}. We emphasize that a point is marked as excluded only if it is excluded for any choice of the remaining free parameters ($f$, $H_I$, $T_b$ and $T_{\mathrm{rh}}$) that is allowed in a given scenario. In particular, this is the reason why the superradiance and the density-induced runaway constraints are stronger in the $T_{\mathrm{rh}} \gg T_b$ scenario, as this scenario restricts the parameter region to smaller masses and, thus, larger decay constants where the constraints are stronger. We do not show the  projected supernova constraints~\cite{Krnjaic:2015mbs, Dev:2020eam, Balaji:2022noj}, as that region is essentially covered by stellar bounds and late decays. Inside the black dashed lines, the relaxion can also explain the DM. We show the contours of the maximal value of the Hubble scale during inflation in this region. We discuss the DM window in the $T_{\mathrm{rh}} \gg T_b$ scenario in the next two subsections.

\subsection{Dark matter from roll-on}
\label{ssec_DMroll}

As it was shown, for small enough $\R$, the relaxion gets re-trapped after reheating. In this section, we construct the DM window for the roll-on misalignment.

\bigskip

\textbf{The CbQ regime:} To reconstruct the window for the CbQ case, we impose the condition from Eq.~(\ref{Hyungjin_get_trapped}) for the relaxion to get retrapped, as it was done in~\cite{Banerjee:2018xmn}. Here the displacement is determined by Eq.~(\ref{Hyungjin_theta}). We also require $T_b < 100 \, \mathrm{GeV}$ for the barrier reappearance temperature. Since the field enters the oscillatory phase much after the barriers reappear i.e.~$H_{\mathrm{osc}}<H_b$, the energy density in the oscillations today can be computed using the standard formula from Eq.~(\ref{eq:dmdensitytd}), inserting (\ref{Hyungjin_theta}) for the field value.

The resulting DM window is shown in the plots of Figs.~\ref{DM1} and~\ref{DM2} using grey color. As can be seen, this window covers different regions of the parameter space as compared to the stochastic window from section~\ref{sec:dm}.

It can be checked that in the parameter region where the above described scenario can explain DM, the relaxion always gets trapped in the first local minimum, even for the largest available value of the Hubble scale during inflation, $H_I^3 = g\Lambda^3$~\cite{Banerjee:2018xmn}. As a consequence the DM window is determined by the values of $g$, $\Lambda$ and $f$ and by physics after inflation, while the value of $H_I$ is irrelevant.

\bigskip

\textbf{The QbC regime:} Larger values of inflationary Hubble scales $H_I$ are available in the QbC regime. It is thus important to find the additional parameter region for the DM window, that opens up if one drops the CbQ condition.

For reasons explained in the previous section, we consider only the case when the field stops at $\Lambda_b ^4 \approx g\Lambda^3f$ and $\delta \ll 1$, i.e.~the QbC I regime. Increasing $H_I$ increases also the stochastic misalignment, which can be computed using Eq.~(\ref{stochastic_1}). To ensure that the stochastic DM component is subdominant we require that the typical misalignment from Eq.~(\ref{stochastic_variance}) is much less compared to the roll-on displacement $\Delta \phi$ from Eq.~(\ref{Hyungjin_theta}). This sets as a new upper bound on $H_I$. The new available parameter region is shown in the lower panel of figure~\ref{DM2}, as the grey shaded region marked as ``QbC''. As expected, it lies above the ``CbQ'' region and even has some overlap with it. We have checked that also inside the additional QbC window the relaxion gets trapped in the first minimum. This explains why the parameter region remains unchanged in the remaining three plots of figures~\ref{DM1} and~\ref{DM2}.

\subsection{Dark matter from stochastic misalignment (QbC)}
\label{ssec_DMstochastic2}

We now return to the stochastic misalignment and demonstrate that a DM window is possible for high reheating temperatures. Here the additional requirement is that the roll-on displacement after inflation is less compared to the stochastic misalignment. In other words, the temperature $T_b$, for which the correct relic density rom roll-on would be generated, should be less than 100 GeV since, otherwise, the roll-on contribution would always dominate. Note that this ensures that the relaxion gets re-trapped and that it does so without changing its local minimum. We also require $T_b>\Lambda_b$. 

The expression for the relic density from Eq.~(\ref{eq:dmdensitytd}) should in principle be modified to cover the case when the barriers appear after $t_{\mathrm{osc}}$ and the onset of oscillations is delayed,
\beq
\frac{ \langle \Omega_{\phi, 0} \rangle  }{ \Omega_{\mathrm{DM}} } \approx  5 \sqrt{ \frac{m_{\phi}}{\mathrm{eV} } } \Bigl( \frac{\sigma_{\phi}}{10^{12}\mathrm{GeV}}\Bigr)^2  \mathcal{F}(T_{\mathrm{osc}}) \times \mathrm{max}\Bigl\{ \Bigl(1, \frac{m_{\phi}}{3H_b}\Bigr)^{3/2}\Bigr\}.
\eeq
As explained in section~\ref{ssec_displacement}, requiring non-eternal inflation implies that the relaxion can get re-trapped only in the QbC I regime, where $\delta \ll 1$ holds. In that case, the barriers necessarily re-appear before the onset of oscillations, thus, there is no delay in the onset of oscillations.

With all formulas at hand, the new DM window can be constructed. It is shown in figures~\ref{DM1} and \ref{DM2} using black solid lines. Compared to the $T_{\mathrm{rh}}<T_b$ case, the parameter region has shrunk. In particular, to derive the new upper bound on the mass for high reheating temperatures, the additional requirement of getting re-trapped should be imposed. The condition (\ref{Hyungjin_get_trapped}) can be re-expressed as $m_{\phi} < 2\sqrt{2} H_b\delta$. The upper bound on $\delta$ can be obtained from the stopping condition from Eq.~(\ref{stopping_condition}), $\delta^3 \approx (9 H_I^4)/{(16\pi^2\Lambda_b^4)}$, inserting $\Lambda_b^4 = g\Lambda^3f$, the upper bound on $H_I$ from (\ref{noneternal}) for non-eternal inflation and the condition from (\ref{Hyungjin_get_trapped}).
One then arrives at the following mass range for relaxion DM
\beq
\label{mass_range_DM_highTrh}
\boxed{ 
10^{-13} \Bigl( \frac{\Lambda}{\mathrm{TeV}} \Bigr)^{\frac{16}{7}}\Bigl( \frac{M_{Pl}}{f} \Bigr)^{\frac{4}{7}} < \frac{m_{\phi}}{ \mathrm{eV}} < 10^{-5}  \Bigl( \frac{g(T_b)}{100} \Bigr) \Bigl( \frac{T_b}{100\mathrm{GeV}} \Bigr)^4 \Bigr( \frac{\mathrm{TeV}}{\Lambda}\Bigl)^2, } 
\eeq 
in the regime of high reheating temperatures. This new range of masses is still larger compared to the case of roll-on misalignment, which can be seen in Fig.~\ref{DM1}. The cut-off scale can be raised to at most $\Lambda \approx 80 \mathrm{TeV}$. The DM windows for the different scenarios are summarized in Fig.~\ref{fig:summary}, where they are shown in black dashed lines.

\section{Thermal production of the relaxion}
\label{ssec:thermal}

In addition to the misalignment mechanism, the relaxions (as well as general ALPs) can be produced in thermal equilibrium, by various processes in the early universe. Here, we consider processes that involve only the couplings of the relaxion due to the Higgs mixing
\cite{Espinosa:2015eda,Flacke:2016szy}. It turns out that, in the regime where the stochastic misliagnment of the relaxion can explain DM, the thermal production is smaller. It is thus only relevant for heavier relaxions.  We sketch the relevant formulas below, following~\cite{Flacke:2016szy}.

At temperatures below the electroweak scale, relaxions can be produced via the Primakoff process, $q(g) + g \rightarrow q(g) + \phi$, and the Compton photoproduction process, $q+g \rightarrow q+\phi$. Here, the first process involves the effective coupling of the Higgs to gluons, while the second one involves the Higgs coupling to quarks. The corresponding production rates are given by~\cite{Masso:2002np, Turner:1986tb}
\beq
\Gamma_{\mathrm{Primakoff}} = 0.3 \frac{\alpha_s^3 \sin^2 \theta_{h\phi} T^3}{\pi^2 v_h^2}, \: \: \: \: \: \: \: \: \Gamma_{\mathrm{Compton}} = \frac{\alpha_s \sin^2 \theta_{h\phi} T \sum_f m^2_f}{\pi^2 v_h^2},
\eeq
where in the second term the dominant contribution comes from the $c$ and $b$ quark.

With the knowledge of the total rate, $\Gamma \approx \Gamma_{\mathrm{Primakoff}} + \Gamma_{\mathrm{Compton}}$, the abundance of thermally produced relaxions can be computed by solving the Boltzmann equation. The solution is~\cite{Flacke:2016szy}
\beq
Y_{\mathrm{thermal}} \approx 0.003\Bigl[1-\exp(-9\times 10^{11} \mathrm{sin}^2\theta_{h \phi} )\Bigr],
\eeq
which is approximately $Y_{\mathrm{thermal}} \approx 0.003$ for $\sin\theta_{h\phi}\gtrsim 10^{-6}$ and $Y_{\mathrm{thermal}} \approx  2.9 \times 10^9 \sin^2\theta_{h\phi}$ for $\sin\theta_{h\phi}\lesssim 10^{-6}$. Here $Y$ denotes the yield, or comoving number density.

From the yield, the relic density can be computed using the following formula,
\beq
\rho_{\phi, 0} = Y m_{\phi} s(T_0) = 0.14 g_{s}(T_0) T_0^3 Y m_{\phi},
\eeq
and the fractional energy density has the form
\beq
\frac{\Omega_{\phi, 0}}{\Omega_{\mathrm{DM}}} = 0.75 \Bigl(\frac{m_{\phi}}{\mathrm{eV}}\Bigr) Y_{\mathrm{thermal}} = 2.4\times 10^{-3} \Bigl(\frac{m_{\phi}}{\mathrm{eV}}\Bigr)   \Bigl[1-\exp(-9\times 10^{11}\mathrm{sin}^2\theta_{h\phi})\Bigr] .
\eeq

Comparing this thermal production with the one from the stochastic misalignment, we find that in the stochastic relaxion DM window, the thermal population is always negligible. Moreover, there is no DM window from the thermal production itself (as also mentioned in~\cite{Flacke:2016szy}) in the region which is not excluded by astrophysical or laboratory probes. 

Thermal production is still relevant to constrain the relaxion parameter space in the region where the relaxion is cosmologically unstable. Depending on the mass of the relaxion, its main decay channels via the Higgs mixing are into a pair of photons, leptons or mesons. The lifetime $\tau_{\phi}$ can be computed as in~\cite{Bezrukov:2009yw, Chatrchyan:2022pcb}. In the region with $1s<\tau_{\phi}<10^{26}s$, which is where the thermal production typically dominates, a number of cosmological constraints apply on the relaxion \cite{Espinosa:2015eda, Flacke:2016szy}. These include the bounds on the baryon-to-photon ratio and on the effective number of neutrino species due to entropy injection, constraints from big bang nucleosynthesis, distortions of the CMB and that of the extragalactic  background light. In Fig.~\ref{fig:summary} these constraints, taken from~\cite{Flacke:2016szy}, are projected into the parameter region of the relaxion in the $[\Lambda, g]$ plane.

\section{The case of the QCD relaxion}
\label{ssec_QCDDM}

\begin{figure}[!t]
\centering
\includegraphics[width=0.48\textwidth]{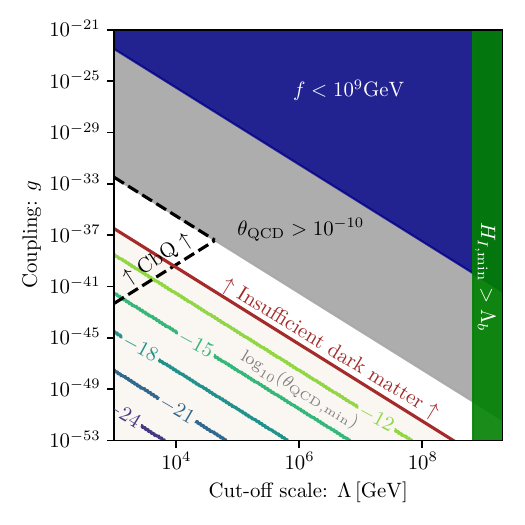}
\includegraphics[width=0.48\textwidth]{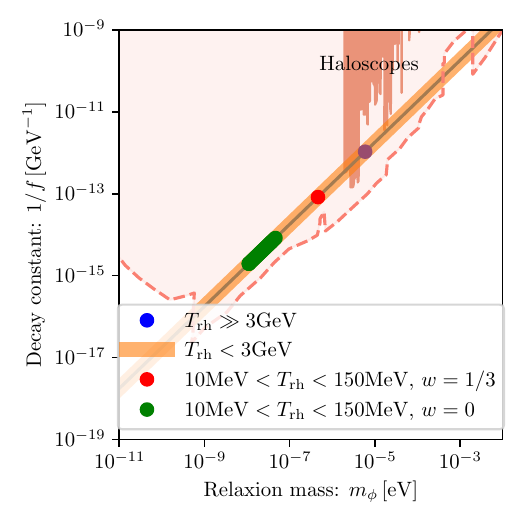}
\includegraphics[width=0.32\textwidth]{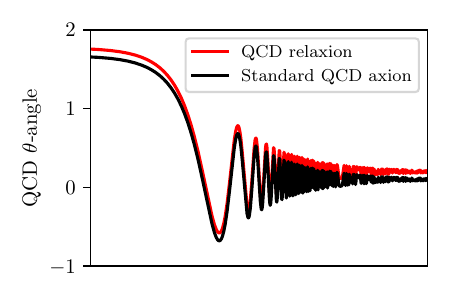}
\includegraphics[width=0.32\textwidth]{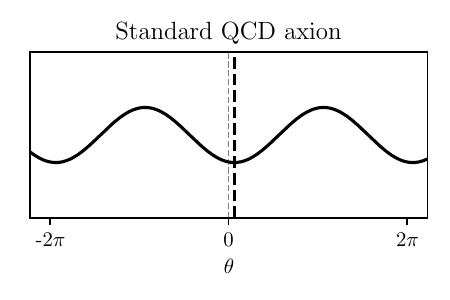}
\includegraphics[width=0.32\textwidth]{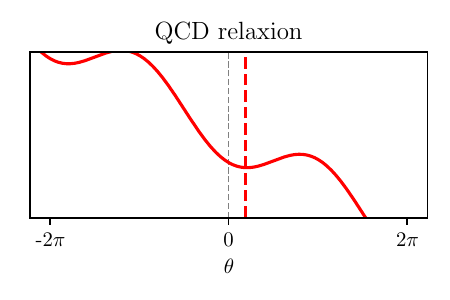}
\caption{Upper row: the QCD relaxion DM window in the $[\Lambda, g]$ (left) and in the $[1/f, m_{\phi}]$ (right) planes. In the left plot, the contours of the minimal value of $\theta_{\mathrm{QCD}}$ are shown inside the region where the relaxion can constitute the totality of DM. In the right plot, the current and projected sensitivities of haloscope experiments are shown, assuming a KSVZ axion model. Different benchmark cases are displayed along the QCD line. Lower row: a schematic illustration of the different value of $\theta_{\mathrm{QCD}}$, determined by the linear slope, in the QCD relaxion model (red), compared to the standard QCD axion case (black) which predicts $\theta_{\mathrm{QCD}}\lesssim 10^{-17}$ (see e.g.~\cite{Georgi:1986kr}). The first panel depicts the decaying oscillations of $\theta_{\mathrm{QCD}}$ while the remaining two panels illustrate the potential energy for both cases.}
\label{DM3}
\end{figure}

The stochastic misalignment of the relaxion in the QbC regime can explain the DM abundance
in the universe. Importantly, this does not require eternal inflation, thus avoids the associated
measure problems~\cite{Gupta:2018wif}.
This is not true however if the relaxion is identified with the QCD
axion, as this case requires eternal inflation, as it was shown in~\cite{Chatrchyan:2022pcb}.
Nevertheless, we still show in this section how the QCD relaxion can be the DM, overlooking the eternal inflation issue.
In the original proposal~\cite{Graham:2015cka}, the QCD axion can be the relaxion in the CbQ regime if a change of the slope of the potential after inflation can be engineered, but the cutoff scale is limited to ${\cal O}$(30) TeV. The corresponding region of parameter space is shown in the upper left plot of figure \ref{DM3} (see also~\cite{Chatrchyan:2022pcb}). In the following, we show that the QCD axion can be the relaxion up to large cutoff scales and  constitute DM. We review both the low and high
reheating temperature cases. The DM discussion is essentially the same in the CbQ and QbC regimes, only the corresponding regions of parameter space are different.

\subsection{QCD relaxion dark matter for $T_{\mathrm{rh}}<T_b$}
The energy density due to  the stochastic misalignment can be estimated using the formulas from section~\ref{ssec:abundace} where, for the QCD relaxion, we require that the reheating temperature does not exceed $T_b = \Lambda_{\mathrm{QCD}} \approx$ 150 {MeV}. The field is typically misaligned by $\Delta \phi \sim f$ from its local minimum, which follows directly from the modified stopping condition $H_I\sim \Lambda_b$~\cite{Chatrchyan:2022pcb}. This is also true for the model of~\cite{Graham:2015cka} where, although $H_I$ is much smaller, the change of the slope of the potential after inflation results in $\Delta \phi \approx (\pi/2)f$. The final slope in both cases should generate only a small CP violating $\theta$-angle,
\beq
\label{theta_angle}
\sin(\theta_{QCD}) = \frac{g\Lambda^3f}{\Lambda_b^4} < 10^{-10},
\eeq
where $\Lambda_b \approx 75 $ MeV. For known values of $T_{\mathrm{rh}}$ and the equation of state of the universe $w$ between inflation and reheating, the relaxion energy density today is determined only by the mass (or the decay constant). The parameter region where the relaxion can explain DM is shown in the upper part of Fig.~\ref{DM3}, as the brown shaded region in the $g$ vs $\Lambda$ plane and as the colored points in the $f^{-1}$ vs $m_{\phi}$ plane. In the second plot, we consider two benchmark cases with $w=0$ (the red point) and $w=1/3$ (the green points). Similar to the standard QCD axion, the relaxion is on the $m_{\phi}f = \Lambda_b^2$ line, shown in blue. Indeed, the expression for the relaxion mass
\beq
m^2_{\phi} = \frac{\Lambda_b^4}{f^2} \sin \delta \approx \frac{\Lambda_b^4}{f^2}\cos(\theta_{QCD}),
\eeq
is very close to the standard expression given that $\theta_{QCD}<10^{-10}$. As can be seen, the cut-off scale for such a DM relaxion can be raised to up to $10^{9} \mathrm{GeV}$.

\subsection{QCD relaxion dark matter for $T_{\mathrm{rh}}\gg T_b$:} Assuming that the universe reheats to temperatures well above $\Lambda_{\mathrm{QCD}}$, the barriers of the potential shrink and the relaxion can roll down. The displacement can be computed using the expression from section~\ref{ssec_displacement} with the only difference being the nontrivial temperature-dependence of the barriers for a QCD axion,
\beq
\label{QCD_T_dep}
\Lambda_b^4(T, h) \approx \frac{\Lambda_b^4(0, h)}{1+(T/\Lambda_{\mathrm{QCD}})^{m}},
\eeq
with $m\approx8.16$~\cite{Borsanyi:2016ksw}. We note that
\begin{itemize}
\item At temperatures $T>\Lambda_{\mathrm{QCD}}\theta_{\mathrm{QCD}}^{-1/m}$ the wiggles are essentially negligible, $\Lambda_b^4(T)<g\Lambda^3 f$, and the solution for the linear potential can be used, $\dot \phi(t) = \frac{2}{5} g\Lambda^3 t [1- (t_{\mathrm{rh}}/t)^{5/2}]$. 

\item For $T_{\mathrm{osc}} < T<\Lambda_{\mathrm{QCD}}\theta_{\mathrm{QCD}}^{-1/m}$, where $\Lambda_b^2(T_{\mathrm{osc}}) = 3H(T_{\mathrm{osc}})f $, the field evolves in a potential with a small slope and its velocity decreases approximately as $a^{-3}$.

\item At temperatures below $T_{\mathrm{osc}}$ the relaxion enters the oscillatory phase. Its mass still increases with time until $T \approx \Lambda_{QCD}$. 
\end{itemize}

The displacement from roll-on at $T_{\mathrm{osc}}$ can be estimated using the above approximations and compared to the stochastic misalignment. As it was already pointed out in~\cite{Graham:2015cka}, for $f>10^{10}$ {GeV}, the first contribution is always very small and, therefore, the stochastic misalignment is unaffected by reheating. The relic density can be computed using Eq.~(\ref{eq:dmdensitytd}), multiplied by a factor $\sqrt{m_{\phi}(T_{\mathrm{osc}})/m_{\phi}}$ due to the temperature-dependence of the mass (see e.g.~\cite{Arias:2012az}). Note that this is the expression for the standard QCD axion DM is shown in the upper part of Fig.~\ref{DM3}.

\bigskip

The main difference compared to the standard QCD axion DM is the possibility to have a large $\theta_{\mathrm{\mathrm{QCD}}}$ angle at the minimum of the relaxion potential. The later is a consequence of the linear term, which shifts the minimum from the CP conserving value according to Eq.~(\ref{theta_angle}). This is illustrated in the bottom part of the figure, where the standard QCD potential is shown in black and the relaxion potential is shown in red. The third bottom plot on the left illustrates how the $\theta_{\mathrm{QCD}}$ parameter is expected to oscillate around the minimum of the potential with a decreasing amplitude due to Hubble expansion. We stress that even in the case of the standard QCD axion CP violation in the SM displaces the minimum of the axion potential from zero. In~\cite{Georgi:1986kr}, the authors estimated $\theta_{\mathrm{QCD}}\sim 10^{-17}$ in the SM (see also~\cite{DiLuzio:2021jfy, Pospelov:2005pr}). The average value of the oscillation amplitude of $\theta_{\mathrm{QCD}}$ today is smaller and approximately
\beq
\bar{\theta}_{\mathrm{QCD}}  \sim \frac{2 \bar{\rho}_{\mathrm{DM}}}{m_{\phi}^2f^2} \approx  \frac{2 \bar{\rho}_{\mathrm{DM}}}{\Lambda_b^4} \sim 10^{-21}.
\eeq
In contrast, in the QCD relaxion model, $\theta_{\mathrm{QCD}}$ oscillates around some value determined by the slope of the potential. In the $g$ vs $\Lambda$ parameter region of Fig.~\ref{DM3} the contours of the minimal values of this angle are shown.

The value of the decay constant $f$ determines the strength of the pseudo-scalar couplings of the axion, and, in particular, the axion-photon coupling. Assuming the KSVZ model where $g_{\phi \gamma \gamma} = \frac{\alpha}{2\pi} \frac{1.92}{f}$, we show the current and projected future sensitivities of haloscopes, taken from~\cite{AxionLimits}, in the $[m_{\phi}, 1/f]$ plane of Fig.~\ref{DM3}.

\section{Summary}
\label{sec:conclusion}

In this work, we have identified a novel scenario in which the original relaxion \cite{Graham:2015cka} naturally constitutes the DM in our universe. 
It is remarkable that the minimal Higgs-axion lagrangian with the simple potential ({\ref{eq_potential}) can address both the Higgs mass hierarchy and the dark matter puzzles. 
The misalignment from the local minimum is generated during the long phase of inflationary dynamics, which is accompanied with a random-walk due to fluctuations. The only requirement here is dropping the CbQ condition for the relaxion dynamics during inflation, which was discussed in detail in our earlier work~\cite{Chatrchyan:2022pcb}. 

This DM scenario from stochastic misalignment  is complementary to the one from~\cite{Banerjee:2018xmn}, where the misalignment of the relaxion originates instead from its  evolution  after inflation. Compared to this latter case, the stochastic DM window covers a wider range of masses and, in particular, allows the relaxion to have larger couplings and smaller decay constants. This is possible both for reheating temperatures not exceeding $T_b < v_h$, as well as in the case of high reheating temperatures. In the first case, the mass range for relaxion DM is given by (\ref{mass_range_DM}) while in the second case it is given by (\ref{mass_range_DM_highTrh}).} The parameter region available for relaxion DM is illustrated in figures~\ref{DMHm},~\ref{DM1},~\ref{DM2} and~\ref{fig:summary}. Among others, figures~\ref{DM1} and~\ref{DM2} show the regions that are excluded by fifth-force experiments, as well as constraints from stellar cooling and black hole superradiance. The parameter region can be further extended and include smaller masses if one allows for the possibility of eternal inflation.

The mixing with the Higgs enables unique search strategies for relaxion DM. In the presence of a coherently oscillating relaxion background, the Higgs mass, hence, most of the fundamental constants of the SM, become oscillatory in time, as explained in~\cite{Banerjee:2018xmn, Banerjee:2019epw}. Such time-variations may potentially be probed by table-top experiments, including atomic, molecular or nuclear clocks~\cite{Arvanitaki:2014faa, Stadnik:2015xbn, Safronova:2017xyt, Aharony:2019iad, Peik:2020cwm}. The strength of oscillations depends on the local DM density and, ignoring the substructure of DM, is beyond the currently projected sensitivity of nuclear clocks~\cite{Banerjee:2020kww}. The sensitivity can be enhanced if a significant fraction of DM is contained in dense localized objects, such as relaxion stars or miniclusters~\cite{Banerjee:2019epw, Banerjee:2019xuy}, which motivates further studies in this direction. We also mention the interesting possibility discussed in~\cite{Banerjee:2019epw, Banerjee:2019xuy}, of a DM overdensity forming around the sun or the earth, which would further enhance the signal. In addition to the Higgs coupling, the relaxion is expected to have pseudo-scalar couplings to the SM and, in that case, can be probed by e.g.~haloscopes.\footnote{We refer to~\cite{AxionLimits, Eroncel:2022vjg} for a detailed list of experiments.}

We have also considered the QCD relaxion model from~\cite{Chatrchyan:2022pcb} and the possibility of such relaxion constituting the DM via its stochastic misalignment. In contrast to the nonQCD models, eternal inflation is required for setting a small value for $\theta_{\mathrm{QCD}}$. The mixing with the Higgs is small and the main interaction channel with the SM is through the pseudo-scalar coupling, which can be probed in haloscope searches. The larger CP violation compared to the standard axion DM scenario, illustrated in Fig.~\ref{DM3}, can be probed by future neutron EDM searches \cite{Abel:2018yeo, Filippone:2018vxf}.

\acknowledgments
	
We are grateful to Hyungjin Kim for many insights as well as collaboration on related work. We also thank Marco Gorghetto, Alessandro Lenoci and Enrico Morgante for useful discussions, as well as Andrea Caputo, Edoardo Vitagliano and Yongchao Zhang for clarifying the stellar bounds.
This work is supported by the Deutsche Forschungsgemeinschaft under Germany Excellence Strategy - EXC 2121 ``Quantum Universe'' - 390833306.

\bibliographystyle{JCAP}
\bibliography{masterbib-new}
	
\end{document}